\documentclass[preprintnumbers]{revtex4}
\usepackage{amsfonts}
\usepackage{amsmath}
\usepackage{graphicx}
\usepackage{dcolumn}
\usepackage{bm}
\usepackage{hyperref}

\input{tcilatex}
\begin{document}

\preprint{}
\title{Optimization and error model for atom interferometry technique to
measure Newtonian gravitational constant}
\author{B. Dubetsky}
\affiliation{bdubetsky@gmail.com}
\date{\today }

\begin{abstract}
Considered contribution to the phase of the atom interferometer caused by
the gravity field of the massive proof mass. Demonstrated the method of
finding the extrema of this contribution for 100kg Tungsten proof mass of
the specific shape and specific parameters of $^{133}Cs$ atom
interferometers. Calculated variations of the double difference response
under the small deviations of atomic and proof mass variables. The choice of
the extremal values of the atomic variables allows one to release
requirements for atom positioning on 2 orders of magnitude.
\end{abstract}

\maketitle
\preprint{}

Atom interference \cite{c1} is one of the tool to measure Newtonian
gravitational constant \cite{c2,c3}. An atomic gravity-gradiometer \cite{c4}
is used in this measurements. When one initially launches the atom cloud \
at position $\vec{a}$ with velocity $\vec{v}$ \ on the one of the hyperfine
sublevel $F_{g}$ of the atomic ground state manifold and\ applies\ at the
moments 
\begin{equation}
\tau =\{t_{1},t_{1}+T,t_{1}+2T\}  \label{1p}
\end{equation}%
$\dfrac{\pi }{2}-\pi -\dfrac{\pi }{2}$ sequence of the Raman pulses resonant
to the atomic transition to another hyperfine sublevel $F_{e},$ the
population of the sublevel $F_{e},$ after interaction, contains \cite{c5}
interferometric term, whose phase is linear on the gravity field $\vec{g}%
\left( \vec{x}\right) .$ In Eq. (\ref{1p}) $t_{1}$ is time delay between
moments of atom launching and first Raman pulse and $T$ is time separation
between pulses. Measuring, in the Earth gravity field $\vec{g}_{E}\left( 
\vec{x}\right) ,$ the phase difference $\Delta \phi $ between two
interferometers with clouds launched at positions and velocities $\left\{ 
\vec{a}_{1},\vec{v}_{1}\right\} $ and $\left\{ \vec{a}_{2},\vec{v}%
_{2}\right\} ,$ one gets signal linear on the Earth gravity field gradient
tensor \cite{c4}. When this gravity-gradiometer operates in the presence of
the proof mass $W,$ the total gravity field 
\begin{equation}
\vec{g}\left( \vec{x}\right) =\vec{g}_{E}\left( \vec{x}\right) +\delta \vec{g%
}\left( \vec{x}\right) ,  \label{2p}
\end{equation}%
where $\delta \vec{g}\left( \vec{x}\right) $ is proof mass gravity field,
and therefore the atom interferometer's phase is linear on $\delta \vec{g}%
\left( \vec{x}\right) .$ Performing these measurements\ for two positions of
the proof mass, which we call below "joined" and "separated" and calculating
the difference of the measurements, one gets the double difference of
phases, $\delta \Delta \phi ,$ which is evidently caused only by the proof
mass field $\delta \vec{g}\left( \vec{x}\right) .$ This double difference of
phase we call response

In this article we determine numerically optimal positions and velocities $%
\left\{ \vec{a}_{i},\vec{v}_{i}\right\} $ to maximize the response and
determine the sensitivity of the response to the variations of the atomic
and proof mass variables. For the part of the atom interferometer phase,
caused by the proof mass, which we call below just "phase", one can use
expression \cite{c6} 
\begin{subequations}
\label{3p}
\begin{align}
\phi & =\vec{k}\cdot \left( \tau _{3}\vec{u}_{30}-t_{1}\vec{u}_{20}+\vec{u}%
_{21}-\vec{u}_{31}\right) ,  \label{3pa} \\
\vec{u}_{\alpha \beta }& =\int_{\tau _{\alpha -1}}^{\tau _{\alpha
}}dtt^{\beta }\delta \vec{g}\left( \vec{a}+\vec{v}t+\vec{g}_{E}\dfrac{t^{2}}{%
2}\right) ,  \label{3pb}
\end{align}%
where $\vec{k}$ is effective Raman wave vector, $\tau _{i}$ is defined in
Eq. (\ref{1p}). Expression (\ref{3p}) was derived under assumptions

\begin{enumerate}
\item proof mass gravity field has small magnitude $\left( \left\vert \delta 
\vec{g}\left( \vec{x}\right) \right\vert \ll \left\vert \vec{g}%
_{E}\right\vert \right) $ but arbitrary inhomogeneity;

\item recoil effect is negligible;

\item Earth gravity field is permanent $\vec{g}_{E}\left( \vec{x}\right)
\equiv \vec{g}_{E}$

\item clouds' temperature is sufficiently small to neglect Raman resonance
Doppler broadening during pulse duration and clouds' thermal expansion
during time $t_{1}+2T;$

\item clouds' size is sufficiently small to neglect ac-Stark shift variation
and wave front curvature along the clouds.
\end{enumerate}

\section{\label{s1}Optimization}

Even though Eq. (3) can be applied for any proof masses, including those
chosen in \cite{c2,c3}, we present here results of former calculations
performed for specific case shown in Fig \ \ref{f1}

\begin{figure}[!t]
\includegraphics[width=11cm]{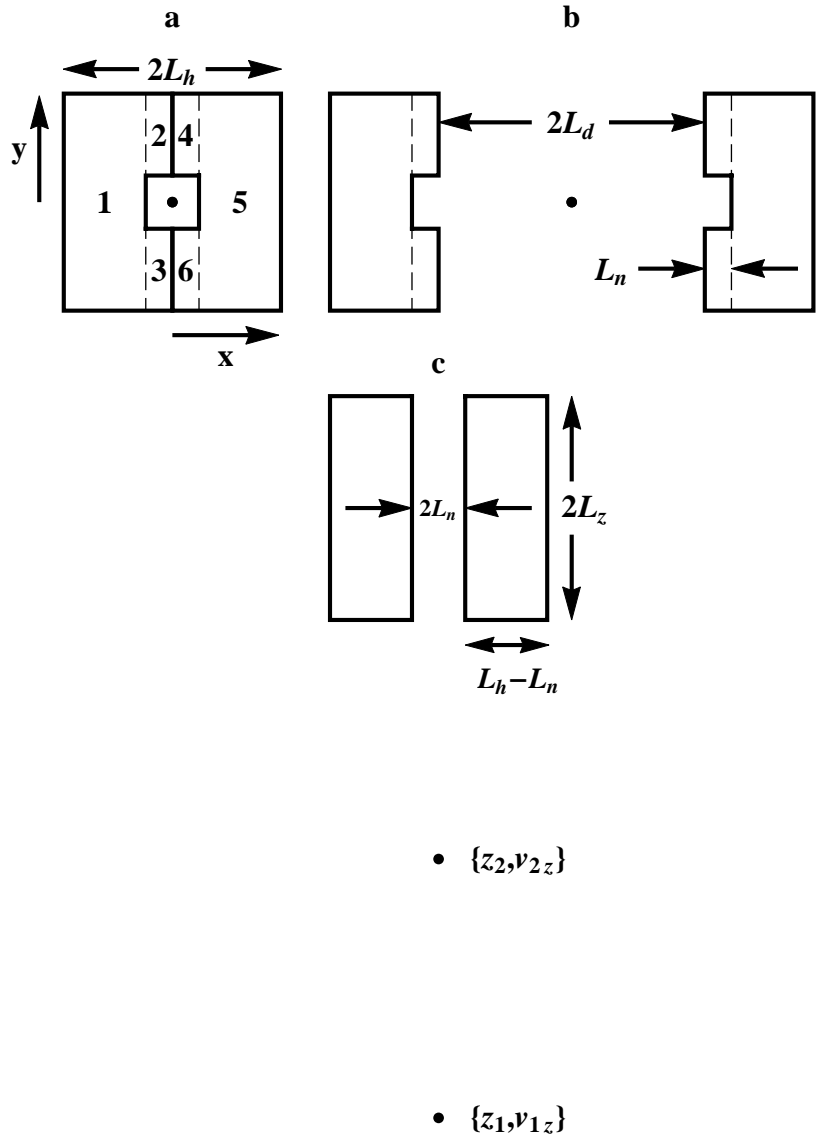}
\caption{The proof mass as a whole is
parallelepiped $2L_{h}\times 2L_{h}\times 2L_{z}$ with narrow $2L_{n}\times
2L_{n}\times 2L_{z}$ hole for Raman fields and atom trajectories. Atoms are
launched vertically from the points $z_{1}$ and $z_{2}$ with velocities $%
v_{1z}$ and $v_{2z}.$ Proof mass consists from 2 halves. (a) Top view.
Joined halves. (b) Top view. Halves separated on the distance $2L_{d}$ along 
$x$ access. (c) Side view, cross-section $x=0.$}
\label{f1}
\end{figure}

Scale of the parameters chosen for calculations are pieced together in the
Table \ref{t1}. The chosen value of density corresponds to pure Tungsten 
\cite{c7}.

\begin{table}[h]
\caption{Order of magnitude of the atom interferometer and proof mass
parameters }%
\begin{tabular}{ll}
Atom & $^{133}Cs$ \\ 
Effective wave vector & $\vec{k}=\left\{ 0,0,k\right\} ,~k=1.47\ast 10^{7}$m$%
^{-1}$ \\ 
Time between launch and first Raman pulse & $t_{1}\sim 40$ms \\ 
Time between Raman pulses & $T\sim 250$ms \\ 
Relative acuracy of atom interferometer phase measurement & $err=10^{-4}$ \\ 
Earth gravity field & $\vec{g}=\left\{ 0,0,-9.8\text{m/s}^{\text{2}}\right\} 
$ \\ 
Proof mass & $W=100$kg \\ 
Proof mass density & $19250$kg/m$^{3}$ \\ 
The hole size & $L_{n}=0.02$m%
\end{tabular}%
\label{t1}
\end{table}

For the given proof mass difference between phases of the interferometers $%
\left\{ z_{1},v_{1z}\right\} $ and $\left\{ z_{2},v_{2z}\right\} $ is
maximal when $\left\{ z_{1},v_{1z}\right\} $ is an absolute maximum of the
phase and $\left\{ z_{2},v_{2z}\right\} $ is an absolute minimum of the
phase. To find out these extrema we used an iterative process, which was
continued until the new value of the extremum differs relatively from the
previous value less than measurement accuracy 
\end{subequations}
\begin{equation}
err=10^{-4}.  \label{4p}
\end{equation}%
Our choice of the proof mass shape is convenient because for the gravity
potential of the parallelepiped having homogeneous density $\rho $ and sizes 
$2a_{x}\times 2a_{y}\times 2a_{z}$ one has analytic expression \cite{c9,c6},%
\begin{eqnarray}
\Phi &=&-G\rho
\sum_{j_{x}=-1}^{1}\sum_{j_{y}=-1}^{1}\sum_{j_{z}=-1}^{1}j_{x}j_{y}j_{z}f%
\left( x+j_{x}a_{x},y+j_{y}a_{y},z+j_{z}a_{z}\right) ,  \notag \\
f\left( u,v,w\right) &=&-\dfrac{1}{2}w^{2}\arctan \left( \dfrac{uv}{wr}%
\right) -\dfrac{1}{2}v^{2}\arctan \left( \dfrac{uw}{vr}\right) -\dfrac{1}{2}%
u^{2}\arctan \left( \dfrac{vw}{ur}\right)  \notag \\
&&+vw\ln \left( u+r\right) +uw\ln \left( v+r\right) +uv\ln \left( w+r\right)
,  \notag \\
r &=&\sqrt{u^{2}+v^{2}+w^{2}}.  \label{5p}
\end{eqnarray}%
We performed calculations for Newtonian gravitational constant $%
G=6.67428\ast 10^{-11}$m$^{3}$s$^{-2}$kg$^{-1}$. The proof mass shown in
Fig. \ref{f1} consists of the parallelepipeds 1,2 and 3 for one half and 4,5
and 6 for another half.

\begin{figure}[!t]
\includegraphics[width=11cm]{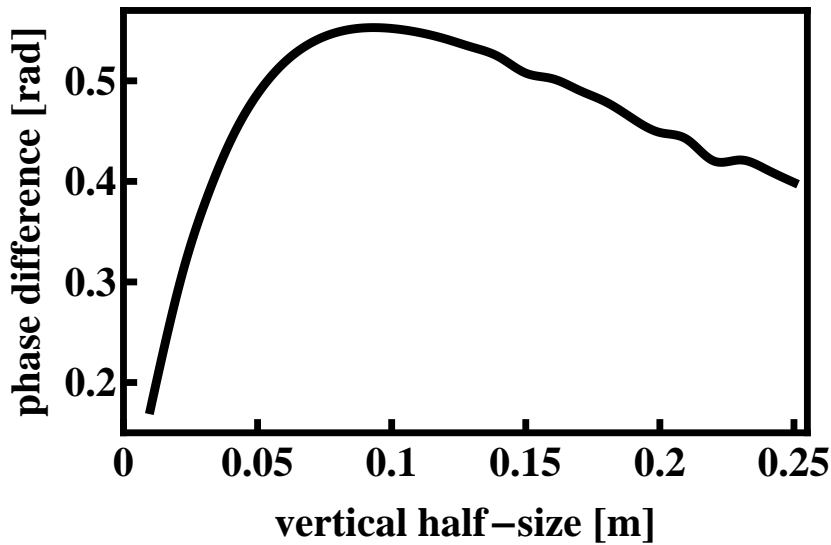}
\caption{Dependence of the maximum of
phase difference on parallelepiped half-size}
\label{f2}
\end{figure}

\begin{figure}[!t]
\includegraphics[width=11cm]{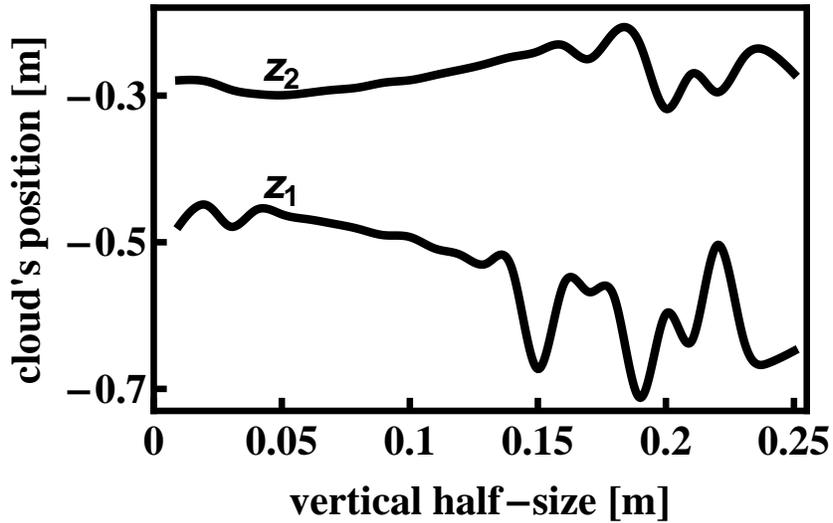}
\caption{Optimal position of the 1st and
2nd atom cloud}
\label{f3}
\end{figure}

\begin{figure}[!t]
\includegraphics[width=11cm]{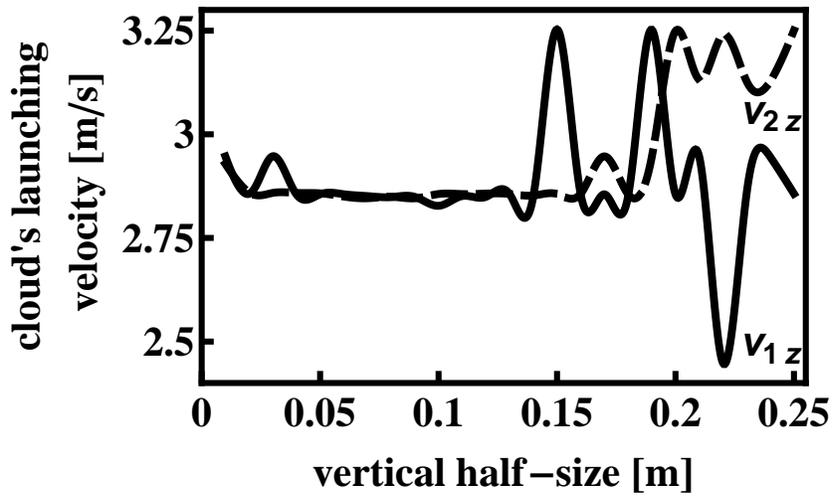}
\caption{Optimal launching velocities of
the 1st and 2nd atom clouds}
\label{f4}
\end{figure}

Dependences of the maximal phase difference, position and velocity of
maximum and minimum on the half-size of the proof mass $2L_{z}$ are shown in
the Figs \ref{f2}-\ref{f4}.

From the Fig. \ref{f2} one sees that the optimal phase difference has its
own maximum. The value of this maximum and values of parameters we recommend
to choose to observe it are presented in Table \ref{t2}.

\begin{table}[h]
\caption{Optimal proof mass sizes and atom clouds positions and velocities
to maximize the phase difference}%
\begin{tabular}{ll}
phase difference & $\Delta \phi =0.55271113$ rad \\ 
vertical half-size & $L_{z}=0.09$m \\ 
horisontal half-size & $L_{h}=0.08726$m \\ 
1st cloud position & $z_{1}=-0.4904$m \\ 
1st cloud launching velocity & $v_{1z}=2.849$m/s \\ 
2nd cloud position & $z_{2}=-0.2823$m \\ 
2nd cloud launching velocity & $v_{2z}=2.846$m/s%
\end{tabular}%
\label{t2}
\end{table}

\section{\label{s2}Error model}

To achieve high precision of the interferometers' phase measurements one has
to prepare with great accuracy both the atomic and proof mass system. In
this section we determine requirements for preparation to achieve phase
measurements with accuracy (\ref{4p}).

The most challenging here is precise positioning of the atom clouds \cite{c3}%
. The preferable here are, evidently, extrema of the clouds position. That
is why found above extrema in $\left\{ z,v\right\} $ space allow one not
only maximize the response, but also make less severe requirements for atom
clouds position, velocity, temperature and size because the response becomes
quadratic on variations of positions and velocities nea extrema.

\begin{figure}[!t]
\includegraphics[width=11cm]{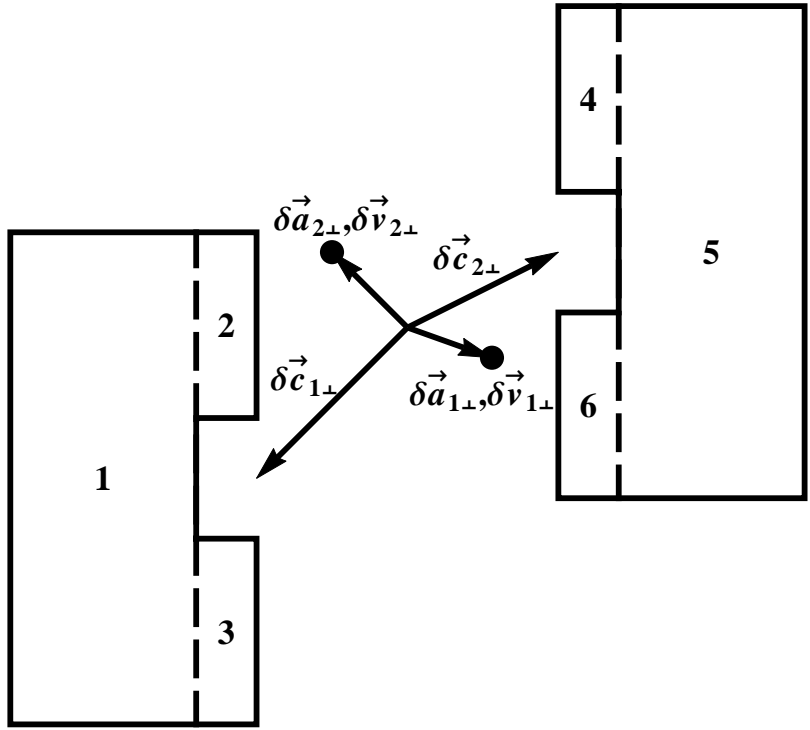}
\caption{Top view. Small variations of
the atomic and proof mass variables. Variations of proof mass halves
orientation and effective wave vector rotation are not shown.}
\label{f5}
\end{figure}

Lets allow now small variations of the atom clouds initial positions,
velocities and effective wave vector (atomic variables) and small
displacement and rotation of proof mass halves (see Fig. \ref{f5}). We
expect that main contribution to response arises from joined proof mass
halves, while for the separated halves contributions to the error decrease
when the distance between halves increases. We determine below, in Sec. \ref%
{s2.1.2.1}, the minimal distance $L_{d},$ starting from which the variations
of contribution to the response from separated halves becomes smaller than
ultimate phase error (\ref{4p}).

\subsection{\label{s2.1}Atomic variables}

\subsubsection{\label{s2.1.1}Joined proof mass halves.}

For the given atom cloud (1 or 2 in Table \ref{t2}) lets denote as $\left\{ 
\vec{a}_{0},\vec{v}_{0}\right\} $ extremal position and velocity and $\vec{k}%
_{0}=\left\{ 0,0,k\right\} ^{T}$ the vertical Raman field effective wave
vector. Including variations one has to substitute in the Eq. (\ref{3p}) 
\begin{subequations}
\label{2}
\begin{align}
\vec{a}& =\vec{a}_{0}+\delta \vec{a},  \label{2a} \\
\vec{v}& =\vec{v}_{0}+\delta \vec{v},  \label{2b} \\
\vec{k}& =R_{\vec{k}}\vec{k}_{0},  \label{2c}
\end{align}%
where $\delta \vec{a}_{i}=\left\{ \delta x_{i},\delta y_{i},\delta
z_{i}\right\} ^{T}$ for interferometer $i.$ We assumed in (\ref{2c}) that
Raman field consists only from counterpropagating wave vectors, but laser
axis could be slightly rotated from direction of $\vec{k}_{0}.$ For the
rotation matrix of this rotation we use Rodriguez rotation formula \cite{c8} 
\end{subequations}
\begin{equation}
R_{\vec{k}ij}=\cos \left( \psi \right) \delta _{ij}+\dfrac{1-\cos \left(
\psi \right) }{\psi ^{2}}\psi _{i}\psi _{j}+\dfrac{\sin \left( \psi \right) 
}{\psi }\varepsilon _{ijm}\psi _{m},  \label{3}
\end{equation}%
where $\vec{\psi}$ is an angle of rotation, $\delta _{ij}$ is Kronecker
symbol, $\varepsilon _{ijm}$ is absolutely antisymmetric tensor. For $\psi
\ll 1,$%
\begin{equation}
R_{\vec{k}ij}\approx \delta _{ij}+\varepsilon _{ijm}\psi _{m}-\dfrac{1}{2}%
\left( \psi ^{2}\delta _{ij}-\psi _{i}\psi _{j}\right) .  \label{4}
\end{equation}

Using this expression and expanding in Eq. (\ref{3p}) up to the 2nd order in
respect to $\delta \vec{a},\delta \vec{v}$ and $\vec{\psi}$ one arrives to
the following approximate expression for the phase

\begin{subequations}
\label{5}
\begin{align}
\phi & \approx \vec{k}_{0}\cdot \left( \tau _{3}\vec{u}_{30}-t_{1}\vec{u}%
_{20}+\vec{u}_{21}-\vec{u}_{31}\right)  \notag \\
& -\left( \vec{\psi}\times \vec{k}_{0}\right) \cdot \left( \tau _{3}\vec{u}%
_{30}-t_{1}\vec{u}_{20}+\vec{u}_{21}-\vec{u}_{31}\right)  \notag \\
& +\delta \vec{a}_{p}\vec{k}_{0i}\left( \tau
_{3}b_{pi30}-t_{1}b_{pi20}+b_{pi21}-b_{pi31}\right)  \notag \\
& +\delta v_{p}\vec{k}_{0i}\left( \tau
_{3}b_{pi31}-t_{1}b_{pi21}+b_{pi22}-b_{pi32}\right)  \notag \\
& +\dfrac{1}{2}\delta \vec{a}_{p}\delta \vec{a}_{q}\vec{k}_{0i}\left( \tau
_{3}d_{pqi30}-t_{1}d_{pqi20}+d_{pqi21}-d_{pqi31}\right)  \notag \\
& +\dfrac{1}{2}\delta v_{p}\delta v_{q}\vec{k}_{0i}\left( \tau
_{3}d_{pqi32}-t_{1}d_{pqi22}+d_{pqi23}-d_{pqi33}\right)  \notag \\
& +\dfrac{1}{2}\left[ \vec{\psi}\times \left( \vec{\psi}\times \vec{k}%
_{0}\right) \right] \cdot \left( \tau _{3}\vec{u}_{30}-t_{1}\vec{u}_{20}+%
\vec{u}_{21}-\vec{u}_{31}\right)  \notag \\
& +\delta \vec{a}_{p}\delta \vec{v}_{q}\vec{k}_{0i}\left( \tau
_{3}d_{pqi31}-t_{1}d_{pqi21}+d_{pqi22}-d_{pqi32}\right)  \notag \\
& -\delta \vec{a}_{p}\left( \vec{\psi}\times \vec{k}_{0}\right) _{i}\left(
\tau _{3}b_{pi30}-t_{1}b_{pi20}+b_{pi21}-b_{pi31}\right)  \notag \\
& -\delta v_{p}\left( \vec{\psi}\times \vec{k}_{0}\right) _{i}\left( \tau
_{3}b_{pi31}-t_{1}b_{pi21}+b_{pi22}-b_{pi32}\right) ,  \label{5a} \\
\vec{u}_{\alpha \beta }& =\int_{\tau _{\alpha -1}}^{\tau _{\alpha
}}dtt^{\beta }\delta \vec{g}\left( \vec{a}_{0}+\vec{v}_{0}t+\vec{g}\dfrac{%
t^{2}}{2}\right) ,  \label{5b} \\
b_{pi\alpha \beta }& \equiv \int_{\tau _{\alpha -1}}^{\tau _{\alpha
}}dtt^{\beta }\partial _{p}\delta \vec{g}_{i}\left( \vec{a}_{0}+\vec{v}_{0}t+%
\vec{g}\dfrac{t^{2}}{2}\right) ,  \label{5c} \\
d_{pqi\alpha \beta }& =\int_{\tau _{\alpha -1}}^{\tau _{\alpha }}dtt^{\beta
}\partial _{p}\partial _{q}\delta \vec{g}_{i}\left( \vec{a}_{0}+\vec{v}_{0}t+%
\vec{g}\dfrac{t^{2}}{2}\right) .  \label{5d}
\end{align}%
A summation convention is implicit in Eq. (\ref{5a}) that will be used in
all subsequent equations, in which repeated indices and symbols are to be
summed over.

We calculated numerically coefficients in the expansion (\ref{5a}) for the
optimal conditions found in Sec. \ref{s1}. Different terms in the Eq. (\ref%
{5a}) are presented in Table \ref{t100}. We changed sign of the terms
associated with interferometer 2.

\begin{table}[h]
\caption{Error model for 100 kg proof mass}%
\begin{tabular}{|c|c|}
\hline
Term & relative weight \\ \hline
Linear in position & 
\begin{tabular}{c}
$-.03159\delta z_{1}$ \\ 
$-0.05332\delta z_{2}$%
\end{tabular}
\\ \hline
Linear in velocity & 
\begin{tabular}{c}
$0.0005170\delta v_{1z}$ \\ 
$-0.0001622\delta v_{2z}$%
\end{tabular}
\\ \hline
nonlinear in position & 
\begin{tabular}{c}
$46.32\left( \delta x_{1}^{2}+\delta y_{1}^{2}\right) $ \\ 
$-92.64\delta z_{1}^{2}$ \\ 
$17.89\left( \delta x_{2}^{2}+\delta y_{2}^{2}\right) $ \\ 
$-35.79\delta z_{2}^{2}$%
\end{tabular}
\\ \hline
nonlinear in velocity & 
\begin{tabular}{c}
$3.981\left( \delta v_{1x}^{2}+\delta v_{1y}^{2}\right) $ \\ 
$-7.962\delta v_{1z}^{2}$ \\ 
$1.587\left( \delta v_{2x}^{2}+\delta v_{2y}^{2}\right) $ \\ 
$-3.173\delta v_{2z}^{2}$%
\end{tabular}
\\ \hline
nonlinear in rotation & 
\begin{tabular}{c}
$-0.2782\left( \psi_{1x}^{2}+\psi_{1y}^{2}\right) $ \\ 
$-0.2218\left( \psi_{2x}^{2}+\psi_{2y}^{2}\right) $%
\end{tabular}
\\ \hline
position-velocity cross term & 
\begin{tabular}{c}
$27.00\left( \delta v_{1x}\delta x_{1}+\delta v_{1y}\delta y_{1}\right) $ \\ 
$-53.99\delta v_{1z}\delta z_{1}$ \\ 
$10.42\left( \delta v_{2x}\delta x_{2}+\delta v_{2y}\delta y_{2}\right) $ \\ 
$-20.83\delta v_{2z}\delta z_{2}$%
\end{tabular}
\\ \hline
position-rotation cross term & 
\begin{tabular}{c}
$-0.01580\left( \delta x_{1}\psi_{1y}-\delta y_{1}\psi_{1x}\right) $ \\ 
$-0.02666\left( \delta x_{2}\psi_{2y}-\delta y_{2}\psi_{2x}\right) $%
\end{tabular}
\\ \hline
velocity-rotation cross term & 
\begin{tabular}{c}
$0.0002585\left( \delta v_{1x}\psi_{1y}-\delta v_{1y}\psi_{1x}\right) $ \\ 
$-0.00008111\left( \delta v_{2x}\psi_{2y}-\delta v_{2y}\psi_{2x}\right) $%
\end{tabular}
\\ \hline
\end{tabular}%
\label{t100}
\end{table}
One sees that in spite of the using extremum points $\left\{
z_{i},v_{i}\right\} $ linear terms are not equal $0.$ It is because extrema $%
\left\{ z_{i},v_{i}\right\} $ have been found in Sec. \ref{s1}
approximately. One can find that coefficients in the linear dependences so
small that for allowed variations of position and velocity (see below Table %
\ref{te100}) linear contributions are negligible.

One can use nonlinear terms to estimate atom clouds' radii and temperatures.
Consider for example relative contribution 
\end{subequations}
\begin{equation}
\delta \varphi _{z}=\alpha \delta z_{i}^{2}.  \label{8}
\end{equation}%
If Raman fields are sufficiently flat to neglect ac-Stark shift variation
across the atom cloud and if Raman pulses are sufficiently short to neglect
the Doppler broadening of the Raman transition, then one needs just to
average (\ref{8}) over atoms' spatial distribution. For Gaussian
distribution, $\dfrac{\exp \left[ -\delta z_{i}^{2}/\delta z_{i\max }^{2}%
\right] }{\sqrt{\pi }\delta z_{i\max }}$, after averaging one gets%
\begin{equation}
\left\langle \delta \varphi _{z}\right\rangle =\dfrac{\alpha }{2}\delta
z_{i\max }^{2}  \label{9}
\end{equation}%
Requiring it to be equal expected relative error of phase measurement, $err,$
one finds for atom cloud radius%
\begin{equation}
\delta z_{i\max }=\sqrt{\dfrac{2\ast err}{\alpha }}.  \label{10}
\end{equation}%
In the same manner we determine atom cloud velocities' variations,
temperatures and angle of the wave vector rotation. These quantities are
pieced together in the Table \ref{te100} for relative error value \ref{4p} 
\begin{table}[h]
\caption{Parameters of the atom interferometers one has to hold for proof
mass 100 kg and relative error $10^{-4}.$}%
\begin{tabular}{cc}
1st cloud vertical radius $\delta z_{1\max }$ [m] & $0.001469$ \\ 
1st cloud vertical velocity $\delta v_{1z\max }$ [m/s] & $0.005012$ \\ 
1st cloud vertical temperature [K] & $2.070\ast 10^{-7}$ \\ 
1st cloud horizontal radius $\delta x_{1\max }$ [m] & $0.002078$ \\ 
1st cloud horizontal velocity $\delta v_{1x\max }$[m/s] & $0.007088$ \\ 
1st cloud horizontal temperature [K] & $4.139\ast 10^{-7}$ \\ 
1st interferometer wave vector rotation angle $\psi _{1\max }$ [rad] & $%
0.02681$ \\ 
2nd cloud vertical radius $\delta z_{2\max }$ [m] & $0.002364$ \\ 
2nd cloud vertical velocity $\delta v_{2z\max }$ [m/s] & $0.007939$ \\ 
2nd cloud vertical temperature [K] & $5.193\ast 10^{-7}$ \\ 
2nd cloud horizontal radius $\delta x_{2\max }$ [m] & $0.003343$ \\ 
2nd cloud horizontal velocity $\delta v_{2x\max }$ [m/s] & $0.01123$ \\ 
2nd cloud horizontal temperature [K] & $1.039\ast 10^{-6}$ \\ 
2nd interferometer wave vector rotation angle $\psi _{2\max }$ [rad] & $%
0.03003$
\end{tabular}%
\label{te100}
\end{table}

\subsubsection{\label{s2.1.2}Separated proof mass halves.}

Contribution to the response from different terms in Eq. (\ref{5a}) arising
for separated proof mass halves are pieced together in the Table \ref{td100}.

\begin{table}[h]
\caption{Contribution to response from separated proof mass halves. Phase
decrease is a ratio of response to the phase difference for joined proof
mass halves. We changed sign for terms related to interferometer 1. Three
values in the curls correspond to the half-distance between proof masses $%
L_{d}=0.15$m, $0.3$m, and $1$m respectively. Values of $\protect\delta %
x_{i\max },\ \protect\delta z_{i\max },\ \protect\delta v_{ix\max },\ 
\protect\delta v_{iz\max },\ \protect\psi _{i\max }$ are taken from table 
\protect\ref{te100}.}%
\begin{tabular}{|c|c|c|}
\hline
Term & relative weight & $%
\begin{array}{c}
\delta x_{i}=\delta y_{i}=\delta x_{i\max },\delta z_{i}=\delta z_{i\max },
\\ 
\delta v_{ix}=\delta v_{iy}=\delta v_{ix\max },\delta v_{iz}=\delta
v_{iz\max },\psi _{i}=\psi _{i\max }%
\end{array}%
$ \\ \hline
Phase decrease & $\left\{ 0.84669512,0.95775894,0.9979968\right\} $ &  \\ 
\hline
\begin{tabular}{c}
Linear \\ 
in position%
\end{tabular}
& 
\begin{tabular}{c}
$\left\{ 0.3435,0.1461,0.009211\right\} \delta z_{1}$ \\ 
$\left\{ -0.5453,-0.1844,-0.009591\right\} \delta z_{2}$%
\end{tabular}
& 
\begin{tabular}{c}
$\left\{ 0.0005046762,0.00021462856,0.000013534148\right\} $ \\ 
$\left\{ -0.0012890354,-0.00043602961,-0.000022672662\right\} $%
\end{tabular}
\\ \hline
\begin{tabular}{c}
Linear \\ 
in velocity%
\end{tabular}
& 
\begin{tabular}{c}
$\left\{ 0.09949,0.04268,0.002719\right\} \delta v_{1z}$ \\ 
$\left\{ -0.1619,-0.05464,-0.002839\right\} \delta v_{2z}$%
\end{tabular}
& 
\begin{tabular}{c}
$\left\{ 0.00049866914,0.00021392743,0.000013627868\right\} $ \\ 
$\left\{ -0.001285252,-0.00043379244,-0.000022539355\right\} $%
\end{tabular}
\\ \hline
\begin{tabular}{c}
nonlinear \\ 
in position%
\end{tabular}
& 
\begin{tabular}{c}
$\left\{ -5.532,-0.8019,-0.006320\right\} \left( \delta x_{1}^{2}+\delta
y_{1}^{2}\right) $ \\ 
$\left\{ 3.751,0.5657,0.004688\right\} \delta z_{1}^{2}$ \\ 
$\left\{ -4.643,-0.6147,-0.004004\right\} \left( \delta x_{2}^{2}+\delta
y_{2}^{2}\right) $ \\ 
$\left\{ 3.241,0.4444,0.002988\right\} $ $\delta z_{2}^{2}$%
\end{tabular}
& 
\begin{tabular}{c}
$\left\{ -4.777\ast 10^{-5},-6.924\ast 10^{-6},-5.458\ast 10^{-8}\right\} $
\\ 
$\left\{ 8.098\ast 10\symbol{94}-6,1.221\ast 10\symbol{94}-6,1.012\ast 10%
\symbol{94}-8\right\} $ \\ 
$\left\{ -0.0001038,-0.00001374,-8.951\ast 10\symbol{94}-8\right\} $ \\ 
$\left\{ 0.00001811,2.484\ast 10\symbol{94}-6,1.670\ast 10\symbol{94}%
-8\right\} $%
\end{tabular}
\\ \hline
\begin{tabular}{c}
nonlinear \\ 
in velocity%
\end{tabular}
& 
\begin{tabular}{c}
$\left\{ -0.5135,-0.07803,-0.0006637\right\} \left( \delta v_{1x}^{2}+\delta
v_{1y}^{2}\right) $ \\ 
$\left\{ 0.3424,0.05444,0.0004912\right\} \delta v_{1z}^{2}$ \\ 
$\left\{ -0.3943,-0.05179,-0.0003348\right\} \left( \delta v_{2x}^{2}+\delta
v_{2y}^{2}\right) $ \\ 
$\left\{ 0.2758,0.03748,0.0002499\right\} $ $\delta v_{2z}^{2}$%
\end{tabular}
& 
\begin{tabular}{c}
$\left\{ -0.00005160,-7.840\ast 10\symbol{94}-6,-6.669\ast 10\symbol{94}%
-8\right\} $ \\ 
$\left\{ 8.602\ast 10\symbol{94}-6,1.367\ast 10\symbol{94}-6,1.234\ast 10%
\symbol{94}-8\right\} $ \\ 
$\left\{ -0.00009940,-0.00001306,-8.442\ast 10\symbol{94}-8\right\} $ \\ 
$\left\{ 0.00003477,4.725\ast 10\symbol{94}-6,3.151\ast 10\symbol{94}%
-8\right\} $%
\end{tabular}
\\ \hline
\begin{tabular}{c}
nonlinear \\ 
in rotation%
\end{tabular}
& 
\begin{tabular}{c}
$\left\{ 0.04398,0.01255,0.0006208\right\} \left( \psi _{1x}^{2}+\psi
_{1y}^{2}\right) $ \\ 
$\left\{ 0.03268,0.008566,0.0003808\right\} \left( \psi _{2x}^{2}+\psi
_{2y}^{2}\right) $%
\end{tabular}
& 
\begin{tabular}{c}
$\left\{ 0.00006323,0.00001805,8.926\ast 10\symbol{94}-7\right\} $ \\ 
$\left\{ 0.00005893,0.00001544,6.867\ast 10\symbol{94}-7\right\} $%
\end{tabular}
\\ \hline
\begin{tabular}{c}
position- \\ 
velocity \\ 
cross term%
\end{tabular}
& 
\begin{tabular}{c}
$\left\{ -3.245,-0.4732,-0.003779\right\} \left( \delta v_{1x}\delta
x_{1}+\delta v_{1y}\delta y_{1}\right) $ \\ 
$\left\{ 2.196,0.3332,0.002802\right\} \delta v_{1z}\delta z_{1}$ \\ 
$\left\{ -2.686,-0.3551,-0.002309\right\} \left( \delta v_{2x}\delta
x_{2}+\delta v_{2y}\delta y_{2}\right) $ \\ 
$\left\{ 1.876,0.2568,0.001724\right\} \delta v_{2z}\delta z_{2}$%
\end{tabular}
& 
\begin{tabular}{c}
$\left\{ -0.00009559,-0.00001394,-1.113\ast 10\symbol{94}-7\right\} $ \\ 
$\left\{ 0.00001617,2.454\ast 10\symbol{94}-6,2.063\ast 10\symbol{94}%
-8\right\} $ \\ 
$\left\{ -0.0002017,-0.00002666,-1.734\ast 10\symbol{94}-7\right\} $ \\ 
$\left\{ 0.00003521,4.819\ast 10\symbol{94}-6,3.235\ast 10\symbol{94}%
-8\right\} $%
\end{tabular}
\\ \hline
\begin{tabular}{c}
position- \\ 
rotation \\ 
cross term%
\end{tabular}
& 
\begin{tabular}{c}
$\left\{ 1.114,0.3560,0.01893\right\} \left( \delta x_{1}\psi _{1y}-\delta
y_{1}\psi _{1x}\right) $ \\ 
$\left\{ -1.434,-0.4119,-0.01944\right\} \left( \delta x_{2}\psi
_{2y}-\delta y_{2}\psi _{2x}\right) $%
\end{tabular}
& 
\begin{tabular}{c}
$0$ \\ 
$0$%
\end{tabular}
\\ \hline
\begin{tabular}{c}
velocity- \\ 
rotation \\ 
cross term%
\end{tabular}
& 
\begin{tabular}{c}
$\left\{ 0.3259,0.1045,0.005593\right\} \left( \delta v_{1x}\psi
_{1y}-\delta v_{1y}\psi _{1x}\right) $ \\ 
$\left\{ -0.4251,-0.1220,-0.005755\right\} \left( \delta v_{2x}\psi
_{2y}-\delta v_{2y}\psi _{2x}\right) $%
\end{tabular}
& 
\begin{tabular}{c}
$0$ \\ 
$0$%
\end{tabular}
\\ \hline
\end{tabular}%
\label{td100}
\end{table}

The point here is that even if the contribution to the response from
separated halves is small this case could be dangerous because launching
positions and velocities found above become no more extrema of the phase,
and therefore major contribution to the phase arises from the linear terms
in Table \ref{td100}. The only way to decrease these linear error is to
increase distance between proof masses $L_{d}$. Indeed for $L_{d}=0.15$m
linear in velocity errors can be 13 times large than ultimate relative
accuracy (\ref{4p}). For $L_{d}=0.3$m they are still 4 times larger. But for 
$L_{d}=1$m all errors linear and nonlinear are well below than parameter $%
err.$

\paragraph{\label{s2.1.2.1}Minimal distance}

\begin{figure}[!t]
\includegraphics[width=11cm]{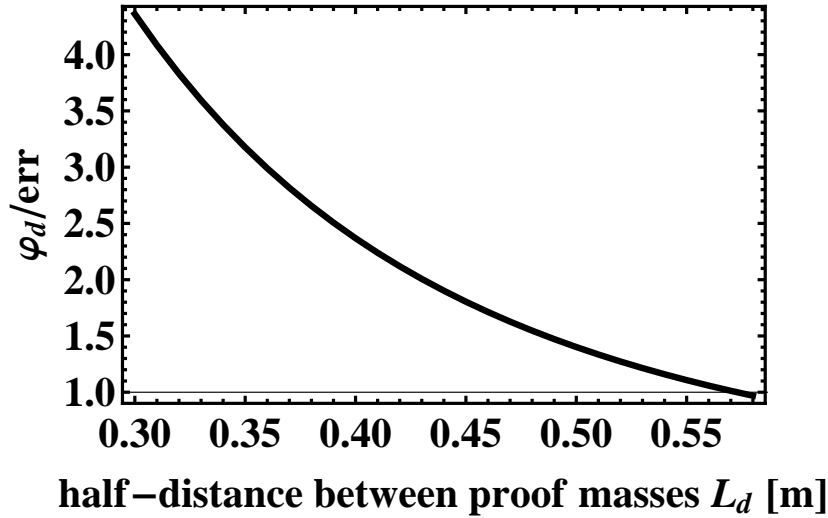}
\caption{Dependence of the error (%
\protect\ref{11.1}) on the half-distance between proof mass halves $L_{d}.$}
\label{f6}
\end{figure}

Since moving proof mass halves on the distances $\pm 1m$ could be a
technological challenge, we determine here minimal half-distance $L_{d}$ of
proof mass halves separation. For quantitative consideration we accept here
that the minimal $L_{d}$ is a distance at which all relative errors in the
3rd columns of the Table \ref{td100} are smaller than parameter $err.$ For
example, for $100$kg proof mass largest error in table \ref{td100} is linear
in position of the second interferometer cloud $\delta z_{2}.$ When
effective wave vector is vertical, $\vec{k}=\left\{ 0,0,k\right\} ,$ from
Eq. (\ref{5a}), one finds for this term%
\begin{equation}
\varphi _{d}\equiv \left\vert k\left( \tau
_{3}b_{3330}-t_{1}b_{3320}+b_{3321}-b_{3331}\right) \right\vert \delta
z_{2\max },  \label{11.1}
\end{equation}%
where tensor $b$ is defined in Eq. (\ref{5c}) and maximal variation of the
atom cloud vertical position, $\delta z_{2\max },$ one finds in the table %
\ref{te100}. Fig. \ref{f6} shows dependence of the term (\ref{11.1}) on the
half distance $L_{d}$.

One sees that $\varphi _{d}$ becomes
smaller than $err$ at $L_{d}=0.58m.$ From the error model for this
half-distance , presented at the table \ref{td100ld} , one sees that all
other errors are also smaller than $err.$%
\begin{table}[tbph]
\caption{The same as in the Table \protect\ref{td100}, but for the distance
between proof mass halves $L_{d}=0.58$m.}%
\begin{tabular}{|c|c|c|}
\hline
Term & relative weight & $%
\begin{array}{c}
\delta x_{i}=\delta y_{i}=\delta x_{i\max },\delta z_{i}=\delta z_{i\max },
\\ 
\delta v_{xi}=\delta v_{yi}=\delta v_{xi\max },\delta v_{zi}=\delta
v_{zi\max },\psi _{i}=\psi _{i\max }%
\end{array}%
$ \\ \hline
Phase decrease & $0.99128250$ &  \\ \hline
\begin{tabular}{c}
Linear \\ 
in position%
\end{tabular}
& 
\begin{tabular}{c}
$0.03715\delta z_{1}$ \\ 
$-0.04099\delta z_{2}$%
\end{tabular}
& 
\begin{tabular}{c}
$0.00005459$ \\ 
$-0.00009691$%
\end{tabular}
\\ \hline
\begin{tabular}{c}
Linear \\ 
in velocity%
\end{tabular}
& 
\begin{tabular}{c}
$0.01093\delta v_{1z}$ \\ 
$-0.01214\delta v_{2z}$%
\end{tabular}
& 
\begin{tabular}{c}
$0.00005477$ \\ 
$-0.00009635$%
\end{tabular}
\\ \hline
\begin{tabular}{c}
nonlinear \\ 
in position%
\end{tabular}
& 
\begin{tabular}{c}
$-0.06795\left( \delta x_{1}^{2}+\delta y_{1}^{2}\right) $ \\ 
$0.04963\delta z_{1}^{2}$ \\ 
$-0.04620\left( \delta x_{2}^{2}+\delta y_{2}^{2}\right) $ \\ 
$0.03420$ $\delta z_{2}^{2}$%
\end{tabular}
& 
\begin{tabular}{c}
$-0.5868\ast 10^{-6}$ \\ 
$0.1071\ast 10^{-6}$ \\ 
$-0.1033\ast 10^{-5}$ \\ 
$0.1911\ast 10^{-6}$%
\end{tabular}
\\ \hline
\begin{tabular}{c}
nonlinear \\ 
in velocity%
\end{tabular}
& 
\begin{tabular}{c}
$-0.006947\left( \delta v_{1x}^{2}+\delta v_{1y}^{2}\right) $ \\ 
$0.005046\delta v_{1z}^{2}$ \\ 
$-0.003872\left( \delta v_{2x}^{2}+\delta v_{2y}^{2}\right) $ \\ 
$0.002867$ $\delta v_{2z}^{2}$%
\end{tabular}
& 
\begin{tabular}{c}
$-0.6980\ast 10^{-6}$ \\ 
$0.1268\ast 10^{-6}$ \\ 
$-0.9761\ast 10^{-6}$ \\ 
$0.1807\ast 10^{-6}$%
\end{tabular}
\\ \hline
\begin{tabular}{c}
nonlinear \\ 
in rotation%
\end{tabular}
& 
\begin{tabular}{c}
$0.002667\left( \psi _{1x}^{2}+\psi _{1y}^{2}\right) $ \\ 
$0.001692\left( \psi _{2x}^{2}+\psi _{2y}^{2}\right) $%
\end{tabular}
& 
\begin{tabular}{c}
$0.3835\ast 10^{-5}$ \\ 
$0.3051\ast 10^{-5}$%
\end{tabular}
\\ \hline
\begin{tabular}{c}
position- \\ 
velocity \\ 
cross term%
\end{tabular}
& 
\begin{tabular}{c}
$-0.04042\left( \delta v_{1x}\delta x_{1}+\delta v_{1y}\delta y_{1}\right) $
\\ 
$0.02949\delta v_{1z}\delta z_{1}$ \\ 
$-0.02666\left( \delta v_{2x}\delta x_{2}+\delta v_{2y}\delta y_{2}\right) $
\\ 
$0.01974\delta v_{2z}\delta z_{2}$%
\end{tabular}
& 
\begin{tabular}{c}
$-0.1191\ast 10^{-5}$ \\ 
$0.2172\ast 10^{-6}$ \\ 
$-0.2001\ast 10^{-5}$ \\ 
$0.3704\ast 10^{-6}$%
\end{tabular}
\\ \hline
\begin{tabular}{c}
position- \\ 
rotation \\ 
cross term%
\end{tabular}
& 
\begin{tabular}{c}
$0.07976\left( \delta x_{1}\psi _{1y}-\delta y_{1}\psi _{1x}\right) $ \\ 
$-0.08506\left( \delta x_{2}\psi _{2y}-\delta y_{2}\psi _{2x}\right) $%
\end{tabular}
& 
\begin{tabular}{c}
$0$ \\ 
$0$%
\end{tabular}
\\ \hline
\begin{tabular}{c}
velocity- \\ 
rotation \\ 
cross term%
\end{tabular}
& 
\begin{tabular}{c}
$0.02351\left( \delta v_{1x}\psi _{1y}-\delta v_{1y}\psi _{1x}\right) $ \\ 
$-0.02518\left( \delta v_{2x}\psi _{2y}-\delta v_{2y}\psi _{2x}\right) $%
\end{tabular}
& 
\begin{tabular}{c}
$0$ \\ 
$0$%
\end{tabular}
\\ \hline
\end{tabular}%
\label{td100ld}
\end{table}

\subsection{\label{s2.2}Proof mass variables}

In this section we consider errors arising from variations of the joined
proof mass halves position and orientation. When the proof mass frame
shifted on $\delta \vec{c}$ and rotated on angle $\vec{\psi}$ in respect to
the lab. frame Eqs. (\ref{3p}) have to be rewritten as 
\begin{subequations}
\begin{align}
\phi & =\vec{k}^{\prime }\cdot \left( \tau _{3}\vec{u}_{30}-t_{1}\vec{u}%
_{20}+\vec{u}_{21}-\vec{u}_{31}\right) ,  \label{12a} \\
\vec{u}_{\alpha \beta }& =\int_{\tau _{\alpha -1}}^{\tau _{\alpha
}}dtt^{\beta }\delta \vec{g}\left( \vec{a}^{\prime }+\vec{v}^{\prime }t+\vec{%
g}_{E}^{\prime }\dfrac{t^{2}}{2}\right) ,  \label{12b}
\end{align}%
where 
\end{subequations}
\begin{subequations}
\label{13}
\begin{align}
\vec{k}^{\prime }& =R\vec{k}_{0},  \label{13a} \\
\vec{a}^{\prime }& =R\left( \vec{a}_{0}-\delta \vec{c}\right) ,  \label{13b}
\\
\vec{v}^{\prime }& =R\vec{v}_{0},  \label{13c} \\
\vec{g}_{E}^{\prime }& =R\vec{g}_{E}  \label{13d}
\end{align}%
are, respectively, wave vector, atoms' launching position, atoms launching
velocity, and Earth gravity field in the proof mass frame, $R$ is rotation
matrix. Configurations considered above could not be optimal for both halves
of the proof mass and we allow the variations of these halves to be
independent then linear in $\delta \vec{c}$ and $\vec{\psi}$ terms should
dominate. So in this section we consider only linear corrections to the
phase, when 
\end{subequations}
\begin{equation}
R_{ij}\approx \delta _{ij}+\varepsilon _{ijm}\psi _{m}.  \label{14}
\end{equation}%
Expanding in Eq. (\ref{12b}) to the linear terms brings one to the following
expression for the phase$\ $%
\begin{align}
\phi & \approx \vec{k}_{0}\left( \tau _{3}\vec{u}_{30}-t_{1}\vec{u}_{20}+%
\vec{u}_{21}-\vec{u}_{31}\right)  \notag \\
& -\left( \vec{\psi}\times \vec{k}_{0}\right) \left( \tau _{3}\vec{u}%
_{30}-t_{1}\vec{u}_{20}+\vec{u}_{21}-\vec{u}_{31}\right)  \notag \\
& -\left( \delta \vec{c}+\vec{\psi}\times \vec{a}_{0}\right) _{p}\vec{k}%
_{i}\left( \tau _{3}b_{pi30}-t_{1}b_{pi20}+b_{pi21}-b_{pi31}\right)  \notag
\\
& -\left( \vec{\psi}\times \vec{v}_{0}\right) _{p}\vec{k}_{i}\left( \tau
_{3}b_{pi31}-t_{1}b_{pi21}+b_{pi22}-b_{pi32}\right)  \notag \\
& -\dfrac{1}{2}\left( \vec{\psi}\times \vec{g}_{E}\right) _{p}\vec{k}%
_{i}\left( \tau _{3}b_{pi32}-t_{1}b_{pi22}+b_{pi23}-b_{pi33}\right) ,
\label{15}
\end{align}%
where tensor $b_{pi\alpha \beta }$ is defined in Eq. (\ref{5c}). For the
chosen proof mas halves' geometry, location and orientation and unperturbed
atomic variables, numeric integration brings one to the following linear
dependence of the phase difference%
\begin{equation}
\Delta \phi \approx 0.5527\left[ 1+6.782\left( \text{$\delta $}c_{1x}-\text{$%
\delta $}c_{2x}\right) +0.04246\left( \text{$\delta $}c_{1z}+\text{$\delta $}%
c_{2z}\right) -0.09261\left( \psi _{1y}-\psi _{2y}\right) \right]  \label{16}
\end{equation}%
where variation of the left (right) half-proof mass position and angle of
rotation are $\delta \vec{c}_{1}$ $\left( \delta \vec{c}_{2}\right) $ and $%
\vec{\psi}_{1}\ \left( \vec{\psi}_{2}\right) .$ One sees that the phase is
most sensitive to displacement along $x-$axis (see Fig. \ref{f1}a). From the
symmetric shapes there are no linear sensitivity to the displacement along $%
y-$axis, rotations in respect to the $z-$ and $x-$axes. When one synchronize
displacement along $x-$axis and rotation of both proof mass halves
corresponding linear dependences disappear. Since in the absence of
rotation\ synchronized displacement of the proof mass halves is equivalent
to the synchronized displacement of both interferometers in the opposite
directions, the slopes of the linear dependences on $\delta z$$_{i}$ equal
to the average slopes in the linear dependences on the interferometers'
displacement taken with opposite sign [compare corresponding coefficients in
Eq. (\ref{16}) and first 2 rows in the Tables \ref{t100}]. Since for optimal
configuration $z-$coordinates of the atom clouds launching points are closed
to the extrema, the slopes of the dependence on $\delta c_{iz}$ in Eq. (\ref%
{16}) are 2 orders of magnitude smaller than slopes of the dependence on $%
\delta c_{ix}.$

From the Eq. (\ref{16}) one concludes that ultimate accuracy (\ref{4p}) can
be achieved for proof mass halves positioning with accuracy 
\begin{subequations}
\begin{eqnarray}
\left\vert \text{$\delta $}c_{ix}\right\vert &<&14.74\mu ,  \label{17a} \\
\left\vert \psi _{iy}\right\vert &<&1.08\text{mrad.}  \label{17b}
\end{eqnarray}

\section{Conclusion}

We showed that $100$kg Tungsten proof mass can produce change in the atom
interferometers phase double difference 
\end{subequations}
\begin{equation}
\delta \Delta \phi =0.54789287\text{rad}  \label{18}
\end{equation}%
for the $^{133}Cs$ atom interferometers, with parameters listed in Table \ref%
{t1}, extremal values of atom interferometers launching position and
velocities and proof mass sizes listed in Table \ref{t2}. The response (\ref%
{18}) is comparable with that observed in \cite{c3}, but the choice of
phase's extrema allowed us to make requirements for atoms' positioning 2
orders of magnitude less severe than requirements for proof mass halves
positioning [compare (\ref{17a}) and data in Table \ref{te100}].

\acknowledgments

Author is appreciated to Drs. M. Kasevich, B. Young, S. Libby, M. Matthews,
T. Loftus, M. Shverdin, V. Sonnad and A. Zorn for fruitful discussion and
collaboration. Special gratefulness to S. Libby and V. Sonnad who showed me
their unpublished results, and to Dr. A. Zorn, who brought to my attention
article \cite{c9}.

The support of this work by Lawrence Livermore National Laboratory, LDRD 12-
LW-009: \textquotedblleft High-Precision Test of the Gravitational
Inverse-Square Law with an Atom Interferometer,\textquotedblright is gratefully acknowledged.
This work was partially performed under the auspices of the U.S. Department
of Energy by Lawrence Livermore National Laboratory under Contract
DE-AC52-07NA27344.


\begin{thebibliography}{9}
\bibitem{c1} B. Dubetsky, A. P. Kazantsev, V. P. Chebotayev, V. P. Yakovlev,
Pis'ma Zh. Eksp. Teor. Fiz. 39, 531 (1984) [JETP Lett. 39, 649 (1984)].

\bibitem{c2} Fixler, J. B., Foster, G. T., McGuirk, J. M. \& M. A. Kasevich,
Science 315, 74 (2007).

\bibitem{c3} G. Rosi, F. Sorrentino, L. Cacciapuoti, M. Prevedelli \& G. M.
Tino, Nature \textbf{510}, 518 (2014).

\bibitem{c4} M. J. Snadden, J. M. McGuirk, P. Bouyer, K. G. Haritos, and M.
A. Kasevich, Phys. Rev. Lett. 81, 971 (1998).

\bibitem{c5} M. Kasevich, S. Chu, Phys. Rev. Lett., 67, 181 (1991).

\bibitem{c6} B. Dubetsky, Private communications, 2008

\bibitem{c7} http://periodictable.com/Properties/A/Density.al.html

\bibitem{c9} Z. F. Seidov, P. I. Skvirsky, arXiv:astro-ph/0002496v1

\bibitem{c8} J. W. Gibbs, E. B. Wilson, "Vector analysis," New York :
Charles Scriber's Sons London: Edward Arnold 1901, p. 338.
\end{thebibliography}
\end{document}